\documentclass[aps,preprint]{revtex4}

\usepackage{amsfonts,amsmath,amssymb}

\begin{document}
\title{ Even-dimensional topological gravity from Chern-Simons gravity.}
\author{Nelson Merino$^{1}$}
\email{nemerino@udec.cl}
\author{Alfredo P\'{e}rez$^{1,2}$}
\email{perez@aei.mpg.de}
\author{Patricio Salgado$^{1}$}
\email{pasalgad@udec.cl}
\date{\today }

\begin{abstract}
It is shown that the topological action for gravity in $2n$-dimensions can be
obtained from the $(2n+1)$-dimensional Chern-Simons gravity genuinely
invariant under the Poincar\'{e} group. The $2n$-dimensional topological
gravity is described by the dynamics of the boundary of a $(2n+1)$-dimensional
Chern-Simons gravity theory with suitable boundary conditions.

The field $\phi^{a},$ which is necessary to construct this type of topological
gravity in even dimensions, is identified with the coset field associated with
the non-linear realizations of the Poincar\'{e} group  $ISO\left(  d-1,1\right)
$ .

\end{abstract}

\address{
$^{1}$Departamento de F\'{\i}sica, Universidad de Concepci\'{o}n, Casilla 160-C, Concepci\'{o}n, Chile\\
\textbf{\textit{$^{2}$}} Max-Planck-Institut f\"{u}r
Gravitationsphysik, Albert-Einstein-Institut, \\
Am M\"{u}hlenberg 1, D-14476 Golm, Germany.}
\maketitle
\section{Introduction}

Twenty years ago A.H Chamseddine \cite{cham1}, \cite{cham2} constructed
topological actions for gravity in all dimensions. Chamseddine showed that the
odd-dimensional theories are based on Chern-Simons forms with the gauge groups
taken to be $ISO(2n,1)$ or $SO(2n+1,1)$ or $SO(2n,2)$ depending on the sign of
the cosmological constant. The use of the Chern-Simons form was essential so
as to have a gauge invariant action without constraints.

The even-dimensional theories use, in addition to the gauge fields, a scalar
multiplet in the fundamental representation of the gauge group. For
even-dimensional spaces there is no natural geometric candidate such as the
Chern-Simons form. The wedge product of $n$ of the field strengths can make
the required $2n$-form in a $2n$-dimensional space-time. The natural gauge
group is $ISO(2n-1,1)$ or $SO(2n,1)$ or $SO(2n-1,2).$ To form a group
invariant $2n$-form, the $n$-product of the field strength is not enough, but
will require in addition a scalar field $\phi^{a}$ in the fundamental representation.

It is the purpose of this paper to show that the topological action for
gravity in $2n$-dimensions can be obtained from the Chern-Simons gravity in
$(2n+1)$-dimensions genuinely invariant under the Poincar\'{e}
group.\ \ \ This Letter is organized as follows: In section $2$ we shall
review some aspects of $(a)$ topological gravity \cite{cham1}, \cite{cham2},
$(b)$ Lanczos-Lovelock gravity theory \cite{lanc}, \cite{lovel}, $(c)$ the
Stelle-West formalism \cite{stelle} and of the Lanczos-Lovelock gravity
theory genuinely invariant under the $AdS$ group \cite{nl1},
\cite{nl2}, \cite{nl3}. In section $3$ it is shown that the topological action
for gravity in $2n$-dimensions, introduced in ref. \cite{cham2} can be
obtained from Chern-Simons gravity in $(2n+1)$-dimensions genuinely invariant
under the Poincar\'{e} group. Section $4$ contains some comments and the conclusions.

\section{\textbf{Review about topological gravity}}

\subsection{\textbf{Actions for topological gravity}}

In Ref. \cite{cham1}, \cite{cham2} A.H Chamseddine constructed topological
actions for gravity in all dimensions. For odd dimensions $d=2n-1,$ the action
is given by \cite{cham1} $S_{2n+1}=k\int_{M_{2n+1}}\omega_{2n+1}$ where
$\omega_{2n+1}$ is a Chern-Simons form given by \cite{naka} $\omega
_{2n+1}=\left(  n+1\right)  \int_{0}^{1}dt\left\langle A\left(  tdA+t^{2}%
A^{2}\right)  ^{n}\right\rangle $ where $A=A^{ab}J_{ab}$ is the Lie algebra
valued one form. Under a gauge transformation the gauge field $A$ transform as
$A^{g}=g^{-1}Ag+g^{-1}dg$ and the Chern-Simons form transforms as \cite{naka}%
\begin{equation}
\omega_{2n+1}^{g}=\omega_{2n+1}+d\alpha_{2n}+(-1)^{n}\frac{n!(n+1)!}%
{(2n+1)!}\left\langle \left(  g^{-1}dg\right)  ^{2n+1}\right\rangle .
\label{cuatro'}%
\end{equation}
Here $\alpha_{2n}$ is a $2n$-form which is a function of $A$.

The even-dimensional theories use, in addition to the gauge fields, a scalar
multiplet in the fundamental representation of the gauge group. To form a
group invariant $2n$-form, the $n$-product of the field strength is not
enough, but will require in addition a scalar field $\phi^{a}$ in the
fundamental representation. \ \ The $2n$-dimensional action is then%
\begin{equation}
I_{2n}=k\int_{M_{2n}}\varepsilon_{aa_{2}\cdot\cdot\cdot\cdot a_{2n+1}\text{ }%
}\phi^{a}F^{a_{2}a_{3}}\cdot\cdot\cdot F^{a_{2n}a_{2n+1}}\text{,
\ }a=0,1,\cdot\cdot\cdot\cdot,2n \label{cinco}%
\end{equation}
where $F^{ab}=dA^{ab}+A^{ac}A_{c}^{\text{ \ }b}.$

This topological gravity has interesting applications, for example in
1+1-dimensions it allows one to describe Liouville's theory for gravity from a
local Lagrangian.

\subsection{\textbf{Lovelock gravity theory}}

In Ref. \cite{zumino}, \cite{teit} was proved that the Lovelock lagrangian
\cite{lanc}, \cite{lovel} can be written as%

\begin{equation}
S=\int\sum_{p=0}^{\left[  d/2\right]  }\alpha_{p}L^{(p)} \label{siete}%
\end{equation}
where $\alpha_{p}$ are arbitrary constants and $L^{(p)}$ is given by
$L^{(p)}=\varepsilon_{a_{1}a_{2}\cdot\cdot\cdot\cdot\cdot\cdot a_{d}}%
R^{a_{1}a_{2}}\cdot\cdot\cdot\cdot R^{a_{2p-1}a_{2p}}e^{a_{2p+1}}\cdot
\cdot\cdot\cdot e^{a_{d}}$ with $R^{ab}=d\omega^{ab}+\omega_{\;c}^{a}%
\omega^{cb}.$

In ref. \cite{tron}\ was shown that requiring that the equations of motion
uniquely determine the dynamics for as many components of the independent
fields as possible fixes the $\alpha_{p}$\ coefficients in terms of the
gravitational and cosmological constants. \ For $d=2n$ the coefficients are
$\alpha_{p}=\alpha_{0}(2\gamma)^{p}\binom{n}{p}$, and the action (\ref{siete})
takes a Born-Infeld-like form. With these coefficients, the LL action is
invariant only under local Lorentz rotations. For $d=2n-1$, the coefficients
become
\begin{equation}
\alpha_{p}=\alpha_{0}\frac{(2n-1)(2\gamma)^{p}}{(2n-2p-1)}\binom{n-1}{p},
\label{cuatro}%
\end{equation}
where $\alpha_{0}=\frac{\kappa}{dl^{d-1}},\qquad\gamma=-$sgn$(\Lambda
)\frac{l^{2}}{2},$ and, for any dimension $d$, $l$ is a length parameter
related to the cosmological constant by $\Lambda=\pm(d-1)(d-2)/2l^{2}.$ With
these coefficients (\ref{cuatro}), the vielbein and the spin connection may be
accommodated into a connection for the AdS group, allowing for the lagrangian
(\ref{siete}) to become the Chern-Simons form in $d=2n+1$ dimensions, whose
exterior derivative is the Euler topological invariant in $d=2n$ dimensions.

\subsection{\textbf{Lovelock gravity theory invariant under Poincar\'{e}
group}}

In Refs. \cite{nl1}, \cite{nl2}, \cite{nl3} it was shown that the Stelle-West
formalism \cite{stelle}, which is an application of the theory of nonlinear
realizations to gravity, permits constructing an action for Lanczos-Lovelock
gravity theory\textbf{\ }genuinely invariant under the $AdS$ group. In fact, a
truly AdS-invariant action for even as well as for odd dimensions was
constructed in ref. \cite{nl2} using the Stelle-West formalism \cite{stelle}%
\ for non-linear gauge theories. The action for this theory is
\begin{equation}
S_{\text{SW}}^{\left(  d\right)  }=\int\sum_{p=0}^{\left[  D/2\right]  }%
\alpha_{p}\varepsilon_{a_{1}\cdots a_{d}}R^{a_{1}a_{2}}\cdots R^{a_{2p-1}%
a_{2p}}V^{a_{2p+1}}\cdots V^{a_{d}}, \label{SSWd}%
\end{equation}
where $V^{a}=\Omega_{\;b}^{a}\left(  \cosh z\right)  e^{b}+\Omega_{\;b}%
^{a}\left(  \frac{\sinh z}{z}\right)  D_{\omega}\phi^{b}$ and $\mathcal{R}%
^{ab}=dW^{ab}+W_{\;c}^{a}W^{cb}$ with $\ W^{ab}=\omega^{ab}+\frac{\sigma
}{l^{2}}\left(  \frac{\sinh z}{z}\right)  e^{c}+\left(  \frac{\cosh z-1}%
{z^{2}}\right)  D_{\omega}\phi^{c}\left(  \phi^{a}\delta_{c}^{b}-\phi
^{b}\delta_{c}^{a}\right)  $ and $\ \Omega_{\;b}^{a}\left(  u\right)  \equiv
u\delta_{b}^{a}+\left(  1-u\right)  \frac{\phi^{a}\phi_{b}}{\phi^{2}}$. Here
$\phi^{a}$ corresponds to the so-called \textquotedblleft(A)dS
coordinate\textquotedblright\ which parametrizes the coset space
$SO_{\hat{\eta}}\left(  D+1\right)  /SO_{\eta}\left(  D\right)  $, and
$z=\phi/l$. This coordinate carries no dynamics, as any value that we pick for
it is equivalent to a gauge choice breaking the symmetry from (A)dS down to
the Lorentz group. This is best seen in the light that the equations of motion
for the action (\ref{SSWd}) are the same as those for ordinary LL gravity,
with $e^{a}$ and $\omega^{ab}$ replaced by $V^{a}$ and $W^{ab}$. The fields
$V^{a}$ and $W^{ab}$ are called non-linear vielbein and spin connection,
respectively, and they take up all the relevant information in the Stelle-West formalism.

From (\ref{SSWd}) we can see that, when one picks the physical gauge $\phi
^{a}=0,$ the theory becomes indistinguishable from the usual one, and the AdS
symmetry is broken down to the Lorentz group. However, a very interesting
exception to this rule occurs in odd dimensions when the coefficients
$\alpha_{p}$ (\ref{cuatro})\ are chosen. In this case, and for any value of
$\phi^{a},$ it is possible to show that the Euler-Chern-Simons action written
with $e^{a}$ and $\omega^{ab}$ differs from that written with $V^{a}$ and
$W^{ab}$ by a boundary term. As a matter of fact, the defining relation for
the non-linear fields $V^{a}$ and $W^{ab}$ given in \cite{stelle}, represents
a gauge transformation for the linear connection $\mathbf{A}=\frac{1}%
{2}i\omega^{ab}\mathbf{J}_{ab}-ie^{a}\mathbf{P}_{a}$, which can be written in
the form $\mathbf{A\rightarrow\tilde{A}}=g^{-1}\left(  d+\mathbf{A}\right)
g,$ where $g=e^{-i\phi^{a}P_{a}}$ and $\mathbf{\tilde{A}}=\frac{1}{2}%
iW^{ab}\mathbf{J}_{ab}-iV^{a}\mathbf{P}_{a}$. This means that the linear and
non-linear curvatures $\mathbf{F}=d\mathbf{A}+\mathbf{A}^{2}$ and
$\mathbf{\tilde{F}}=d\mathbf{\tilde{A}}+\mathbf{\tilde{A}}^{2}$ are related by
$\mathbf{\tilde{F}}=g^{-1}\mathbf{F}g.$ Just as the usual Euler-Chern-Simons
lagrangian, the odd-dimensional non-linear lagrangian, with the special choice
of coefficients given in eq. (\ref{cuatro}), satisfies $dL_{VW}^{\left(
2n-1\right)  }=\left\langle \mathbf{\tilde{F}}^{n}\right\rangle ,$ where
$\left\langle \mathbf{J}_{a_{1}a_{2}}\cdots\mathbf{J}_{a_{2n-3}a_{2n-2}%
}\mathbf{P}_{a_{2n-1}}\right\rangle =\frac{1}{l}\varepsilon_{a_{1}\cdots
a_{2n-1}}$. This implies that
\begin{equation}
dL_{VW}^{\left(  2n-1\right)  }=\left\langle \mathbf{\tilde{F}}^{n}%
\right\rangle =\left\langle \mathbf{F}^{n}\right\rangle =dL_{e\omega}^{\left(
2n-1\right)  },
\end{equation}
and hence we see that both lagrangians may locally differ only by a total
derivative. The same arguments lead to the conclusion that, in general, any
Chern-Simons lagrangian written with non-linear fields, which is genuinelly
invariant, differs from the usual one by a total derivative.

\bigskip It is direct to show that in the limit $l\longrightarrow\infty$ we
obtain a Lovelock gravity theory genuinelly invariant under the Poincar\'{e}
group:%
\begin{equation}
S_{SW}^{\left(  d\right)  }=k\int\varepsilon_{a_{1}\cdots a_{d}}R^{a_{1}a_{2}%
}\cdots R^{a_{d-2}a_{d-1}}V^{a_{d}},
\end{equation}
where now $V^{a}=e^{a}+D_{\omega}\phi^{a}=d\phi^{a}+\omega_{\text{ \ }b}%
^{a}\phi^{b}+e^{a}$ and $R^{ab}=d\omega^{ab}+\omega_{\;c}^{a}\omega^{cb},$
with $W^{ab}=\omega^{ab},$ and $\phi^{a}$ corresponds to the so-called
\textquotedblleft Poincar\'{e} coordinate\textquotedblright. \ The fields
$\phi^{a},e^{a},\omega^{ab}$ under local Poincar\'{e} translations change as
$\delta\phi^{a}=-\rho^{a};$ \ $\delta e^{a}=\kappa_{\text{ \ }b}^{a}e^{b};$
\ $\delta\omega^{ab}=-D\kappa^{ab}.$

\section{\textbf{Topological gravity from Chern-Simons gravity}}

In this section we show that the topological action for gravity in
$2n$-dimensions \cite{cham2} can be obtained from (2n+1)--dimensional
Chern-Simons gravity genuinelly invariant under the Poincar\'{e} group.

The Lanczos-Lovelock action genuinelly invariant under the Poincar\'{e} group
is given by
\begin{equation}
S=k\int\epsilon_{a_{1}....a_{2n+1}}R^{a_{1}a_{2}}...R^{a_{2n-1}a_{2n}%
}V^{a_{2n+1}}. \label{gt9}%
\end{equation}

Introducing the non-linear gauge fields $V^{a}$ into (\ref{gt9}) we obtain%
\[
S=k\int_{M}\epsilon_{a_{1}....a_{2n+1}}R^{a_{1}a_{2}}...R^{a_{2n-1}a_{2n}%
}e^{a_{2n+1}}%
\]%
\begin{equation}
+k\int_{M}\epsilon_{a_{1}....a_{2n+1}}R^{a_{1}a_{2}}...R^{a_{2n-1}a_{2n}}%
D\phi^{a_{2n+1}}, \label{gt9'}%
\end{equation}%
\[
S=k\int_{M}\epsilon_{a_{1}....a_{2n+1}}R^{a_{1}a_{2}}...R^{a_{2n-1}a_{2n}%
}e^{a_{2n+1}}%
\]%
\begin{equation}
+k\int_{M}d\left[  \epsilon_{a_{1}....a_{2n+1}}R^{a_{1}a_{2}}...R^{a_{2n-1}%
a_{2n}}\phi^{a_{2n+1}}\right]  , \label{gt10'}%
\end{equation}
where we have used the Bianchi identity $DR^{ab}=0.$ So that%
\[
S=k\int_{M}\epsilon_{a_{1}....a_{2n+1}}R^{a_{1}a_{2}}...R^{a_{2n-1}a_{2n}%
}e^{a_{2n+1}}%
\]%
\begin{equation}
+k\int_{\partial M}\epsilon_{a_{1}....a_{2n+1}}R^{a_{1}a_{2}}...R^{a_{2n-1}%
a_{2n}}\phi^{a_{2n+1}}. \label{gt11}%
\end{equation}

This action differs by a boundary term from the usual \ Chern-Simons action
for the Poincar\'{e} group written in terms of linear fields. This extra
boundary term allows the action (\ref{gt9}) to be genuinely gauge invariant
and not only modulo boundary terms like the usual action in terms of linear
gauge fields. This extra boundary term also changes the dynamic behavior of
the Chern-Simons theory at the boundary of the manifold as we show below.

From (\ref{gt11}) we can see that the whole dependence of the coset field
$\phi^{a}$ is on the surface term and that the form of the surface term
exactly coincides with the form of the even-dimensional action for
even-dimensional topological gravity. \ However, the surface term cannot be
directly considered as an action principle for the boundary, because the
dynamics of the boundary is determined by the dynamics of the Bulk. We will
show that, for solutions with suitable boundary conditions, it is possible to
obtain the dynamics for the even-dimensional topological action. It must be
noticed that the coset field $\phi^{a}$, which is associated with a non-linear
realization of the Poincar\'{e} group, appears in the action (\ref{gt11}) in a
geometrically natural form.

\subsection{\textbf{Invariance of the action}}

We show now that the action (\ref{gt11}) is invariant under local Poincar\'{e}
translations. In fact, under the transformations%
\begin{equation}
\delta e^{a}=-D\rho^{a};\text{ \ }\delta\phi^{a}=\rho^{a} \label{gt12}%
\end{equation}
we obtain%
\[
\delta_{trasl}S=-k\int_{M}\epsilon_{a_{1}....a_{2n+1}}R^{a_{1}a_{2}%
}...R^{a_{2n-1}a_{2n}}D\rho^{a_{2n+1}}%
\]%
\begin{equation}
+k\int_{\partial M}\epsilon_{a_{1}....a_{2n+1}}R^{a_{1}a_{2}}...R^{a_{2n-1}%
a_{2n}}\rho^{a_{2n+1}} \label{gt12'}%
\end{equation}%
\[
=-k\int_{M}d\left[  \epsilon_{a_{1}....a_{2n+1}}R^{a_{1}a_{2}}...R^{a_{2n-1}%
a_{2n}}\rho^{a_{2n+1}}\right]
\]%
\begin{equation}
+k\int_{\partial M}\epsilon_{a_{1}....a_{2n+1}}R^{a_{1}a_{2}}...R^{a_{2n-1}%
a_{2n}}\rho^{a_{2n+1}}, \label{gt13}%
\end{equation}
where we have used the Bianchi identity $DR^{ab}=0.$ This means that%
\[
\delta_{trasl}S=-k\int_{\partial M}\epsilon_{a_{1}....a_{2n+1}}R^{a_{1}a_{2}%
}...R^{a_{2n-1}a_{2n}}\rho^{a_{2n+1}}%
\]%
\begin{equation}
+k\int_{\partial M}\epsilon_{a_{1}....a_{2n+1}}R^{a_{1}a_{2}}...R^{a_{2n-1}%
a_{2n}}\rho^{a_{2n+1}}=0. \label{gt14}%
\end{equation}

\subsection{\textbf{Equations of motion}}

The variations of the action (\ref{gt11}) with respect to $e^{a},\omega
^{ab},\phi^{a}$ lead to%
\[
\delta S=k\int_{M}n\epsilon_{a_{1}....a_{2n+1}}D\left(  \delta\omega
^{a_{1}a_{2}}\right)  R^{a_{3}a_{4}}...R^{a_{2n-1}a_{2n}}e^{a_{2n+1}}%
\]%
\[
+\epsilon_{a_{1}....a_{2n+1}}R^{a_{1}a_{2}}...R^{a_{2n-1}a_{2n}}\delta
e^{a_{2n+1}}%
\]%
\[
+k\int_{\partial M}n\epsilon_{a_{1}....a_{2n+1}}D\left(  \delta\omega
^{a_{1}a_{2}}\right)  R^{a_{3}a_{4}}...R^{a_{2n-1}a_{2n}}\phi^{a_{2n+1}}%
\]%
\begin{equation}
+\epsilon_{a_{1}....a_{2n+1}}R^{a_{1}a_{2}}...R^{a_{2n-1}a_{2n}}\delta
\phi^{a_{2n+1}}=0. \label{gt15}%
\end{equation}

\bigskip\ Integrating by parts we have%
\[
\delta S=k\int_{M}n\epsilon_{a_{1}....a_{2n+1}}\delta\omega^{a_{1}a_{2}%
}R^{a_{3}a_{4}}...R^{a_{2n-1}a_{2n}}T^{a_{2n+1}}%
\]%
\[
+\epsilon_{a_{1}....a_{2n+1}}R^{a_{1}a_{2}}...R^{a_{2n-1}a_{2n}}\delta
e^{a_{2n+1}}%
\]%
\[
+k\int_{\partial M}n\epsilon_{a_{1}....a_{2n+1}}\delta\omega^{a_{1}a_{2}%
}R^{a_{3}a_{4}}...R^{a_{2n-1}a_{2n}}e^{a_{2n+1}}%
\]%
\[
+n\epsilon_{a_{1}....a_{2n+1}}\delta\omega^{a_{1}a_{2}}R^{a_{3}a_{4}%
}...R^{a_{2n-1}a_{2n}}D\phi^{a_{2n+1}}%
\]%
\begin{equation}
+k\int_{\partial M}\epsilon_{a_{1}....a_{2n+1}}R^{a_{1}a_{2}}...R^{a_{2n-1}%
a_{2n}}\delta\phi^{a_{2n+1}}=0. \label{gt16}%
\end{equation}

From (\ref{gt16}) we can see that by imposing the boundary conditions
\[
\int_{\partial M}n\epsilon_{a_{1}....a_{2n+1}}\delta\omega^{a_{1}a_{2}%
}R^{a_{3}a_{4}}...R^{a_{2n-1}a_{2n}}\left[  e^{a_{2n+1}}+D\phi^{a_{2n+1}%
}\right]
\]%
\begin{equation}
+\int_{\partial M}\epsilon_{a_{1}....a_{2n+1}}R^{a_{1}a_{2}}...R^{a_{2n-1}%
a_{2n}}\delta\phi^{a_{2n+1}}=0 \label{gt17}%
\end{equation}
we obtain the following movement equations:%

\begin{equation}
\epsilon_{a_{1}....a_{2n+1}}R^{a_{1}a_{2}}...R^{a_{2n-1}a_{2n}}=0 \label{gt18}%
\end{equation}%
\begin{equation}
\epsilon_{a_{1}....a_{2n+1}}R^{a_{1}a_{2}}...R^{a_{2n-3}a_{2n-2}}T^{a_{2n-1}%
}=0, \label{gt19}%
\end{equation}
which correspond to the usual Chern-Simons field equations.

From (\ref{gt17}) we can see that the boundary condition associated with
$\delta\phi^{a}$ is identically satisfied due to the validity of equation
(\ref{gt18}) on the boundary. \ Let's now consider a boundary condition of the
Neumann type for the spin connection and, for the vielbein, a boundary
condition of the type $e^{a}\mid_{\partial M}=0$. \ This means that any
solution of the field equations (\ref{gt18}), (\ref{gt19})\ has a void
vielbein on the boundary. \ Introducing the boundary condition $e^{a}%
\mid_{\partial M}=0$ into (\ref{gt17}) we obtain the following condition:%

\begin{equation}
\epsilon_{a_{1}....a_{2n+1}}R^{a_{3}a_{4}}...R^{a_{2n-1}a_{2n}}D\phi
^{a_{2n+1}}\mid_{\partial M}=0 \label{gt21}%
\end{equation}
Therefore, the dynamics of the boundary will be characterized by the following
set of equations:
\begin{align}
\epsilon_{a_{1}....a_{2n+1}}R^{a_{3}a_{4}}...R^{a_{2n-1}a_{2n}}D\phi
^{a_{2n+1}}  &  \mid_{\partial M}=0\label{gt22}\\
\epsilon_{a_{1}....a_{2n+1}}R^{a_{1}a_{2}}...R^{a_{2n-1}a_{2n}}  &
\mid_{\partial M}=0 \label{gt23}%
\end{align}
which can be obtained from the action principle%
\begin{equation}
S^{\left(  2n\right)  }=k\int_{\partial M_{2n+1}=M_{2n}}\epsilon
_{a_{1}....a_{2n+1}}R^{a_{1}a_{2}}...R^{a_{2n-1}a_{2n}}\phi^{a_{2n+1}}.
\label{gt24}%
\end{equation}
This action correspond to topological gravity of ref.\cite{cham2}.

The condition $e^{a}\mid_{\partial M}=0$ can be imposed due to the fact that
$e^{a}$ is a part of a connection, namely $A=\frac{i}{2}\omega^{ab}%
J_{ab}-ie^{a}P_{a}$. To have vielbeins and therefore non-invertible metrics in
the boundary of a manifolf means that the boundary of a manifold defines a
singularity in the metric sector of the theory. However, from the point of
view of the gauge structure, this does not represent any singularity due to
the fact that the vielbein can be annulled because it is part of a gauge connection.

This shows that configurations with non invertible vielbeins can play an
important role in the structure of the theory. Configurations with
singularities on the boundary are not new (ref. \cite{grigna}). They
correspond to natural configurations of the gauge theories for gravity.

Finally it is interesting to notice that now the geometric origin of the
$\phi^{a}$ field is clear, due to the fact that this one is a coset field
associated with non-linear realizations of the Poincar\'{e} group.

\section{\textbf{Comments}}

We have shown in this work that the topological action for gravity in
$2n$-dimensions, introduced in ref. \cite{cham2}, \ can be obtained from the
Chern-Simons gravity in $(2n+1)$-dimensions genuinely invariant under the
Poincar\'{e} group. The $2n$-dimensional topological gravity is described by
the dynamics of the boundary of the $(2n+1)$ Chern-Simons gravity theory with
suitable boundary conditions. \ The boundary of the manifold defines a
singular hypersurface in the metric sector of the theory. The singularity
appears only when we consider configurations with metric invertibles. However,
this singularity is not an intrinsic singularity of the theory due to the fact
that the vielbein is a part of a gauge connection.

The dynamics on the boundary of a $(2n+1)$-dimensional manifold is described
by the field equations of the topological gravity of ref. \cite{cham2}.

The field $\phi^{a}$, which is necessary to construct this type of topological
gravity in even dimensions \cite{cham2}, is identified by the coset field
associated with non-linear realizations of the Poincare group $ISO(2n,1)$.
This shows a clear geometric interpretation of this field originally
introduced "ad-hoc".

\begin{acknowledgments}
This work was supported in part by Direcci\'{o}n de Investigaci\'{o}n,
Universidad de Concepci\'{o}n through Grant \# 208.011.048-1.0 and in part by
FONDECYT through Grants \#s 1080530 and 1070306 . One of the authors (N.M) was
supported by grants from the Comisi\'{o}n Nacional de Investigaci\'{o}n
Cient\'{\i}fica y Tecnol\'{o}gica CONICYT and from the Universidad de
Concepci\'{o}n, Chile.

One of the authors (A.P) wishes to thank to S. Theisen for his kind
hospitality at the M.P.I f\"{u}r Gravitationsphysik in Golm where part of this
work was done. He is also grateful to German Academic Exchange Service (DAAD)
and Consejo Nacional de Ciencia y Tecnolog\'{\i}a (CONICYT) for financial support.
\end{acknowledgments}

\end{document}